\documentstyle[proc,epsf,rotate]{rspublic}
\def\citename{}
\input epsf

\def\thedemobiblio#1{\smallskip\par
 \list{}{\labelwidth 0pt \leftmargin 1em \itemindent -1em \itemsep 1pt}
 \small \parindent 0pt
 \parskip 1.5pt plus .1pt\relax
 \def\newblock{\hskip .11em plus .33em minus .07em}
 \sloppy\clubpenalty4000\widowpenalty4000
 \sfcode`\.=1000\relax}

\def\jrbref{\item}
\def\gs{\mathrel{\lower0.6ex\hbox{$\buildrel {\textstyle >}
 \over {\scriptstyle \sim}$}}}
\def\ls{\mathrel{\lower0.6ex\hbox{$\buildrel {\textstyle <}
 \over {\scriptstyle \sim}$}}}
\newcount\equationo
 at 10truept
 at 10truept

\def\figcapsize{\small \baselineskip=0.01cm}
\def\mathrelfun#1#2{\lower3.6pt\vbox{\baselineskip0pt\lineskip.9pt
  \ialign{$\mathsurround=0pt#1\hfil##\hfil$\crcr#2\crcr\sim\crcr}}}

\def\hmpc{{\, {\rm h}^{-1}~\rm Mpc}}

\def\kms{{\rm~km~s^{-1}}}

\def\mpc{{\rm~Mpc}}

\def\'{^{\prime}}
\def\avrg#1{{\langle #1 \rangle}}

\def\hq{{\hat q}}

\def\eps{\varepsilon}

\def\eg{{e.g., }}
\def\ie{{i.e., }}
\def\etal{{et al. }}
 
\def\etc{{etc. }}

\def\half{{\textstyle{1\over2}}}

\def\p3m{P$^3$M}

\def\spose#1{\hbox to 0pt{#1\hss}}
\def\lta{\mathrel{\spose{\lower 3pt\hbox{$\mathchar"218$}}
     \raise 2.0pt\hbox{$\mathchar"13C$}}}
\def\gta{\mathrel{\spose{\lower 3pt\hbox{$\mathchar"218$}}
     \raise 2.0pt\hbox{$\mathchar"13E$}}}
\def\ge{\mathrel{\spose{\lower 3pt\hbox{$-$}}
     \raise 2.0pt\hbox{$\mathchar"13E$}}}
\def\le{\mathrel{\spose{\lower 3pt\hbox{$-$}}
     \raise 2.0pt\hbox{$\mathchar"13C$}}}

\begin{document}

\title[Constraining LSS with the CMB]{Constraining Large Scale Structure Theories\\
with the Cosmic Background Radiation }

\author[J.R. Bond \& A.H. Jaffe]{J. Richard Bond$^{1}$ and Andrew H. Jaffe$^{2}$}

\affiliation{1. CIAR Cosmology Program, Canadian Institute
for Theoretical Astrophysics, \\
60 St. George St., Toronto, ON M5S 3H8, Canada\\
2. Center for Particle Astrophysics, UC Berkeley, Berkeley CA 94720, USA} 

\maketitle

\abstract 
\noindent
The case is strong that cosmic microwave background (CMB) and large
scale structure (LSS) observations can be combined to determine the
theory of structure formation and the cosmological parameters that
define it. We review: the relevant 10+ parameters associated with the
inflation model of fluctuation generation and the matter content of
the Universe; the relation between LSS and primary and secondary CMB
anisotropy probes as a function of wavenumber; how COBE constraints on
energy injection rule out explosions as a dominant source of LSS; and
how current anisotropy band-powers in multipole-space, at levels $\sim
(10^{-5})^2$, strongly support the gravitational instability theory
and suggest the universe could not have reionized too early. We use
Bayesian analysis methods to determine what current CMB and CMB+LSS
data imply for inflation-based Gaussian fluctuations in tilted
$\Lambda$CDM, $\Lambda$hCDM and $o$CDM model sequences with
cosmological age 11-15 Gyr, consisting of mixtures of baryons, cold
``c'' (and possibly hot ``h'') dark matter, vacuum energy
``$\Lambda$'', and curvature energy ``o'' in open cosmologies. For
example, we find the slope of the initial spectrum is within about 5\%
of the (preferred) scale invariant form when just the CMB data is
used, and for $\Lambda$CDM when LSS data is combined with CMB; with
both, a nonzero value of $\Omega_\Lambda$ is strongly preferred
($\approx 2/3$ for a 13 Gyr sequence, similar to the value from
SNIa). The $o$CDM sequence prefers $\Omega_{tot}<1 $, but is overall
much less likely than the flat $\Omega_\Lambda \ne 0$ sequence with
CMB+LSS. We also review the rosy forecasts of angular power spectra
and parameter estimates from future balloon and satellite experiments
when foreground and systematic effects are ignored to show where
cosmic parameter determination can go with just CMB information alone.
\endabstract

\section{The Relation Between CMB and LSS Observables}\label{sec:theory}

In this section, we first present an overview of the relation between
the scales that CMB anisotropies probe, those that large scale
structure observations of galaxy clustering probe, and the scales that
are responsible for collapsed object formation in hierarchical models
of structure formation, in particular those determining the abundances
of clusters and galaxies. We review the basic parameters of amplitude
and tilt characterizing the fluctuations in the simplest versions of
inflation, but consider progressively more baroque inflation models
needing progressively more functional freedom in describing
post-inflation fluctuation spectra. We then describe the high
precision that has been achieved in calculations of primary CMB
anisotropies (those determinable with linear perturbation theory), and
the less precisely calculable secondary anisotropies arising from
nonlinear processes in the medium. 

\subsection{CMB as a Probe of Early Universe Physics} \label{secCMBprobe}

The source of fluctuations to input into the cosmic structure
formation problem is likely to be found in early universe physics. We
want to measure the CMB (and large scale structure) response to these
initial fluctuations. The goal is to peer into the
physical mechanism by which the fluctuations were generated. The
contenders for generation mechanism are (1) ``zero
point'' quantum noise in scalar and tensor fields that must be there
in the early universe if quantum mechanics is applicable and (2)
topological defects which may arise in the inevitable phase
transitions expected in the early universe.  

From CMB and LSS observations we hope to learn: the statistics of the
fluctuations, whether Gaussian or non-Gaussian; the mode, whether
adiabatic or isocurvature scalar perturbations, and whether there is a
significant component in gravitational wave tensor perturbations; the
power spectra for these modes, $ {\cal P}_{\Phi}(k), {\cal P}_{is}(k)
, {\cal P}_{GW}(k)$ as a function of comoving wavenumber $k=
2\pi\bar{a}/\lambda$, with the cosmological scale factor $\bar{a}(t)$
removed from the physical wavelength $\lambda (t)$ and set to unity
now. The length unit is $\hmpc$, where ${\rm h}$ is the Hubble
parameter in units of $100 \kms \mpc^{-1}$, {\it i.e.}, really a
velocity unit. Until a few years ago ${\rm h}$ was considered to be
uncertain by a factor of two or so, but is now thought to be between
0.6 and 0.7.

Sample initial and evolved power spectra for the gravitational
potential $ {\cal P}_{\Phi}(k)$ ($ \equiv d\sigma_\Phi^2 /d\ln k$, the
{\it rms} power per $d\ln k$ band) are shown in
Fig.~\ref{fig:probes}. The (linear) density power spectra, ${\cal
P}_\rho (k) \propto k^4{\cal P}_{\Phi}(k)$, are also shown in
Fig.~\ref{fig:probes}. (We use ${\cal P}(k)=k^3 P(k)/(2\pi^2)$ for
power spectra, the variance in the fluctuation variable per $\ln k$,
rather than the oft-plotted mean-squared fluctuation for mode $k$, $
P(k)$, so ${\cal P}_\rho \equiv \Delta^2 (k)$ in the notation of
Peacock (1997).) As the Universe evolves the initial shape of ${\cal
P}_{\Phi}$ (nearly flat or scale invariant) is modified by
characteristic scales imprinted on it that reflect the values of
cosmological parameters such as the energy densities of baryons, cold
and hot dark matter, in the vacuum (cosmological constant), and in
curvature.  Many observables can be expressed as weighted integrals
over $k$ of the power spectra and thus can probe both density
parameters and initial fluctuation parameters.

\subsection{Cosmic Structure and the Nonlinear Wavenumber} \label{seckNL}

\begin{figure}
\begin{center}
\epsfysize=5.0in\leavevmode\epsffile{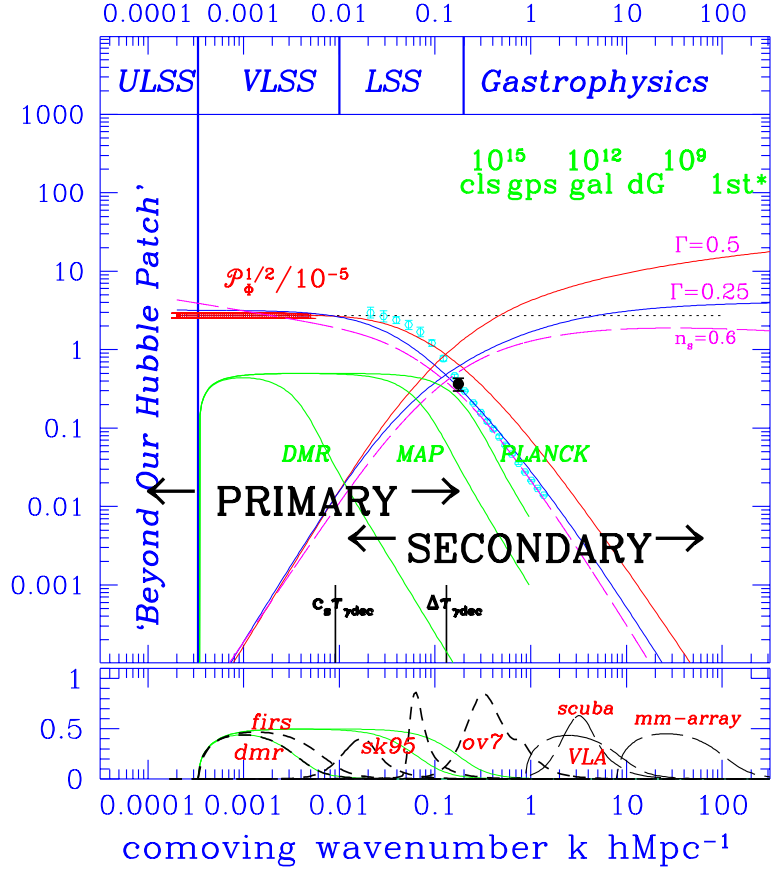}
\end{center}
\caption{\figcapsize The bands in comoving wavenumber $k$ probed by
CMB primary and secondary anisotropy experiments, in particular by the
satellites COBE, MAP and Planck, and by large scale structure (LSS)
observations are contrasted. The width of the CMB photon decoupling
region and the sound crossing radius ($\Delta \tau_{\gamma dec},
c_s\tau_{\gamma dec}$) define the effective acoustic peak range for
primary anistropies (those involving linear fluctuations). Secondary
anisotropies arise only once matter has gone nonlinear.  Sample
(linear) gravitational potential power spectra (actually ${\cal
P}_\Phi^{1/2}(k)$) are also plotted, and the $y$-axis values refer to
${\cal P}_\Phi^{1/2}/10^{-5}$ (which is dimensionless).  The
horizontal dotted line is the post-inflation scale invariant power
spectrum, which is bent down as the universe evolves by an amount
dependent upon the matter content.  The hatched region at low $k$
gives the 4 year DMR error bar on the $\Phi$ amplitude in the COBE
regime.  The solid data point in the cluster-band denotes the $\Phi$
constraint from the abundance of clusters (for
$\Omega_{tot}$=1,$\Omega_{\Lambda}$=0). The open circles are estimates
of the linear $\Phi$ power from current galaxy clustering data by
Peacock (1997). A bias is ``allowed'' to (uniformly) raise the shapes
to match the observations.  The corresponding linear density power
spectra, ${\cal P}_\rho^{1/2}(k)$, are also shown rising to high
$k$. Models shown in Fig.~\ref{fig:probes} are the ``standard''
$n_s=1$ CDM model (labelled $\Gamma=0.5$ with $\Omega_{nr}=1$, ${\rm
h}=0.5$), a tilted ($n_s=0.6$, $\Gamma=0.5$) CDM model and a model
with the shape modified ($\Gamma=0.25$) by changing the matter content
of the Universe, {\it e.g.}, $\Omega_{nr}=0.36$, ${\rm h=0.7}$.  The
bands at high $k$ associated with object formation (cls, gal, \etc)
and the filters showing the bands various CMB experiments probe are
discussed in the text.  }
\label{fig:probes} 
\end{figure}

 In hierarchical structure formation models such as those considered
here, as the universe evolves ${\cal P}_\rho^{1/2}(k)$ grows with
time until it crosses unity at small scales, and the first star
forming tiny dwarf galaxies appear (``1st *''), typically at a redshift
of about 20. The nonlinear wavenumber $k_{NL}(t)$, defined by
$\int_0^{k_{NL}} {\cal P}_\rho (k)d\ln k$=1, decreases as the universe
expands, leaving in its wake dwarf galaxies (dG), normal galaxies
(gal), groups (gps) and clusters (cls), forming from waves
concentrated in the $k$-space bands that their labels cover in
Fig.~\ref{fig:probes}. Equivalent mass scales are given above them.

Scales just below $k_{NL}$ are weakly nonlinear and define the
characteristic patterns of filaments connecting clusters, and
membranes connecting filaments. Voids are rare density minima which
have opened up by gravitational dynamics and merged, opposite to the
equally abundant rare density maxima, the clusters, in which the space
collapses by factors of 5-10 and more. 

At $k >k_{NL}(t)$, nonlinearities and complications associated with
dissipative gas processes can obscure the direct connection to the
early universe physics. Most easily interpretable are observables
probing the linear regime now, $k < k_{NL}(t_0)$. CMB anisotropies
arising from the linear regime are termed primary.  As
Fig.~\ref{fig:probes} shows, these probe three decades in wavenumber, with
the high $k$ cutoff defined by the physics at $z\sim 1000$ when CMB
photons decoupled, not $k_{NL}$ at that time. Within the LSS band,
two important scales for the CMB arise: the sound crossing distance at
photon decoupling, $\sim 100 \hmpc$, and the width of the region over
which this decoupling occurs, which is about a factor of 10 smaller,
and below which the primary CMB anisotropies are damped. LSS
observations of galaxy clustering at low redshift probe a smaller
range, but which overlaps the CMB range. We have hope that $z\sim 3$
LSS observations, when $k_{NL}(t)$ was larger, can extend the range,
but gas dynamics can modify the relation between observable and power
spectrum in complex ways. Although probes based on catalogues of high
redshift galaxies and quasars, and on quasar absorption lines from the
intergalactic medium, represent a very exciting observational
frontier, it will be difficult for theoretical conclusions about the
early universe and the underlying fluctuations to be divorced from
these ``gastrophysical'' complications. Secondary anisotropies of the CMB
(\S~\ref{secsec}), those associated with nonlinear phenomena, also
probe smaller scales and the ``gastrophysical'' realm.

\subsection{Probing Wavenumber Bands with the CMB and LSS} \label{seckprobe}

Although the scales we can probe most effectively are smaller than the
size of our Hubble patch ($\sim 3000 \hmpc$), because ultralong waves
contribute gentle gradients to CMB observables, we can in fact place
useful constraints on the ultralarge scale structure (ULSS) realm
``beyond our horizon''. Indeed current constraints on the size of the
universe arise partly from this region and partly from the very large
scale structure (VLSS) region. (For compact spatial manifolds, the
wavenumbers have an initially discrete spectrum, and are missing
ultralong waves, limited by the size of the manifold.)

The COBE data and CMB experiments with somewhat higher resolution
probe the VLSS region very well. Density fluctuations are highly
linear in that regime, which is what makes it so simple to
analyze. One of the most interesting realms is the LSS one, in which
CMB observations probe exactly the scales that LSS redshift surveys
probe. The density fluctuations are linear to weakly nonlinear in this
realm, so we can still interpret the LSS observations reasonably well
--- with one important caveat: Galaxies form and shine through complex
nonlinear dissipative processes, so how they are distributed may be
rather different than how the total mass is distributed.  The evidence
so far is consistent with this ``bias'' being only a linear amplifier
of the mass fluctuations on large scales, albeit a different one for
different galaxy types.  Detailed comparison of the very large CMB and
LSS redshift survey results we will get over the next five years
should help enormously in determining the statistical nature of the
bias.

Because the ${\cal P}_{\Phi}$ of the COBE-normalized sCDM model shown
shoots high relative to the cluster data point, the sCDM model is
strongly ruled out. More rigorous discussion of what is compatible
with COBE, smaller angle CMB experiments such as SK95, the cluster
data point and the shape of the ${\cal P}_{\Phi}$ spectrum as
estimated from galaxy clustering data is given in
\S~\ref{seccmbLSScurrent}. The filter functions plotted for SK95,
Planck, \etc show the bands they are sensitive to: multiplying by a
$k$-space $\Delta T/T$ power spectrum gives the variance per $\ln k$
(\eg Bond 1996).

\subsection{The Cosmic Parameters of Structure Formation Theories} 
\label{seccosmicparam}

Even simple Gaussian inflation-generated fluctuations for structure
formation have a large number of early universe parameters we would
wish to determine (\S~\ref{secinflation}): power spectrum amplitudes
at some normalization wavenumber $k_n$ for the modes present, $\{
{\cal P}_{\Phi}(k_n), {\cal P}_{is}(k_n) , {\cal P}_{GW}(k_n) \}$;
shape functions for the ``tilts'' $\{ \nu_s(k),\nu_{is}(k), \nu_t(k)
\} $, usually chosen to be constant or with a logarithmic correction,
\eg $\nu_s(k_n), d\nu_s(k_n)/d\ln k$. (The scalar tilt for adiabatic
fluctuations, $\nu_s (k) \equiv d\ln {\cal P}_{\Phi}/d \ln k$, is
related to the usual index, $n_s$, by $\nu_s = n_s-1$.)  The transport
problem (\S~\ref{sectransport}) is dependent upon physical processes,
and hence on physical parameters. A partial list includes the Hubble
parameter ${\rm h}$, various mean energy densities $\{ \Omega_{tot},
\Omega_B , \Omega_{\Lambda}, \Omega_{cdm }, \Omega_{hdm }\}{\rm h}^2$,
and parameters characterizing the ionization history of the Universe,
\eg the Compton optical depth $\tau_C$ from a reheating redshift
$z_{reh}$ to the present. Instead of $\Omega_{tot}$, we prefer to use
the curvature energy parameter, $\Omega_k\equiv 1-\Omega_{tot}$, thus
zero for the flat case. In this space, the Hubble parameter, ${\rm h}=
(\sum_j (\Omega_j{\rm h}^2 ))^{1/2}$, and the age of the Universe,
$t_0$, are functions of the $\Omega_j{\rm h}^2$. The density in
nonrelativistic (clustering) particles is
$\Omega_{nr}=\Omega_B+\Omega_{cdm}+\Omega_{hdm}$.
\footnote{It is becoming conventional to refer to $\Omega_{nr}$ as
$\Omega_m$.} The density in relativistic particles, $\Omega_{er}$,
includes photons, relativistic neutrinos and decaying particle
products, if any.  $\Omega_{er}$, the abundance of primordial helium,
\etc should also be considered as parameters to be determined. The
count is thus at least 17, and many more if we do not restrict the
shape of ${\cal P}_{\Phi}(k)$ through theoretical considerations of
what is ``likely'' in inflation models. Estimates of errors on a
smaller 9 parameter inflation set for the MAP and Planck satellites
are given in \S~\ref{seccmbfuture}.

\begin{figure}[htbp]
\begin{center}
\epsfysize=6.0in\leavevmode\epsffile{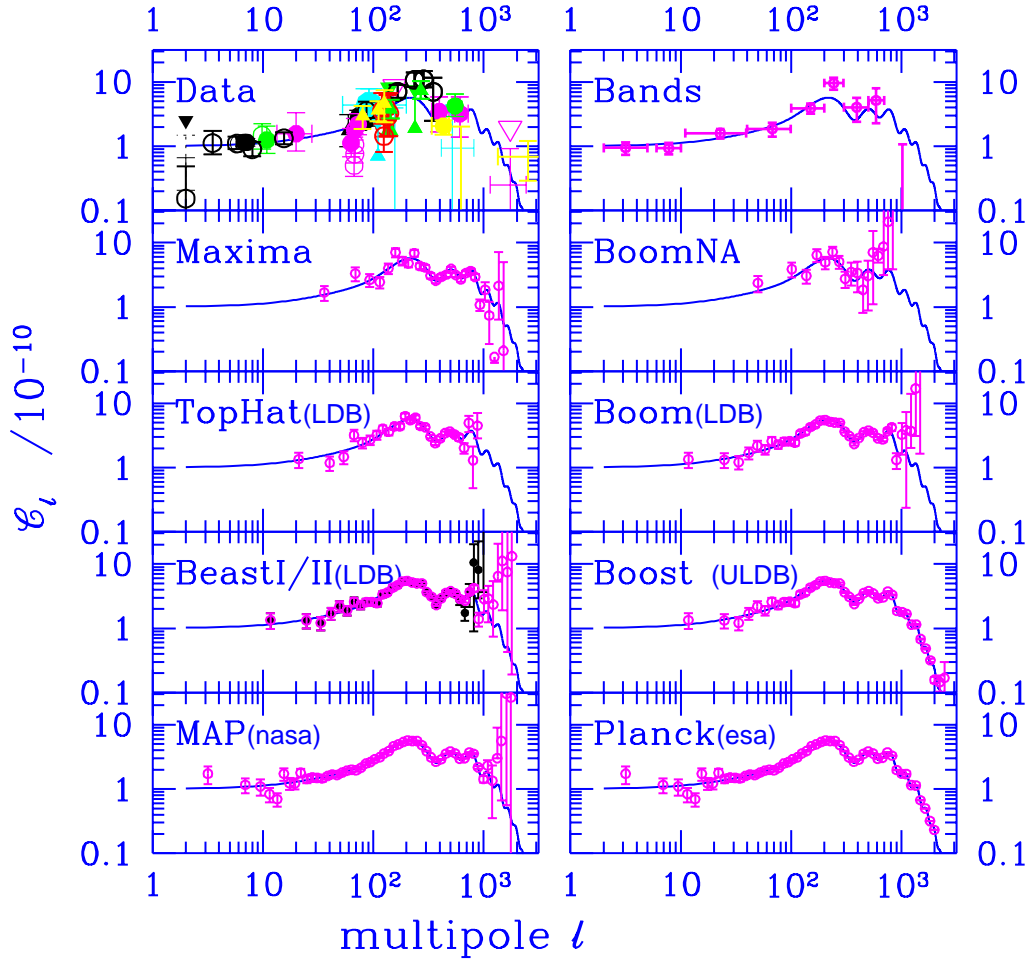}
\end{center}
\caption{\figcapsize The ${\cal C}_\ell$ anisotropy bandpower data for
experiments up to summer 1998 are shown in the upper left panel. The
data are optimally combined into 9 bandpower estimates (with one-sigma
errors) shown in the upper right panel. To guide the eye an untilted
COBE-normalized sCDM model is repeated in all panels. The rest of the
panels show forecasts of how accurate ${\cal C}_\ell$ will be
determined for this model from balloon and satellite experiments, with
parameters given in Table~\ref{tab:exptparams}. }
\label{fig:CLdat}
\end{figure}

\begin{figure}[htbp]
\begin{center}
\epsfysize=6.0in\leavevmode\epsffile{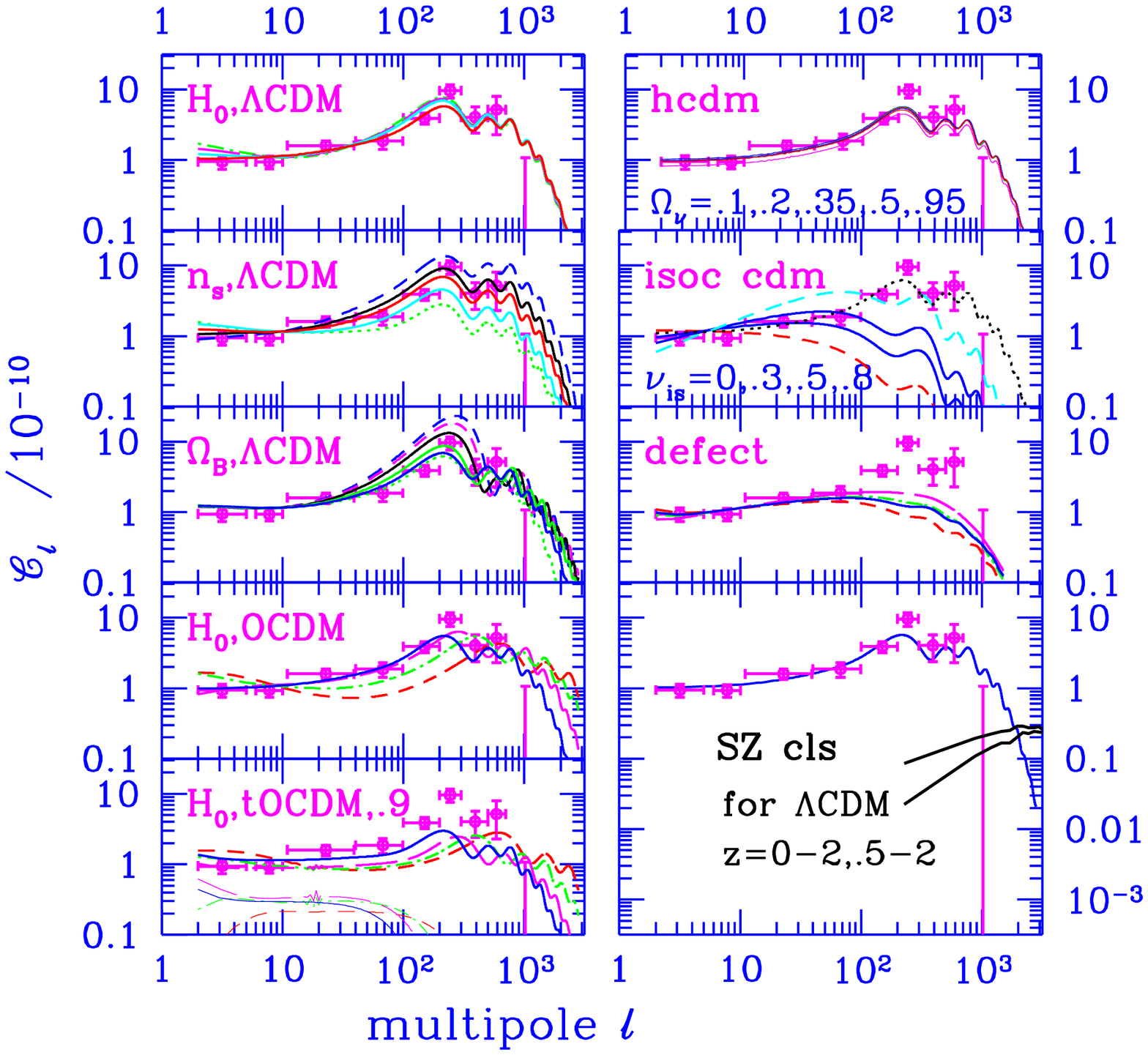}
\end{center}
\caption{\figcapsize  
The 9 bandpower estimates from current anisotropy data are
compared with various 13 Gyr model sequences: (1) $H_0$ from 50 to 90,
$\Omega_{\Lambda}$, 0 to 0.87, for an untilted $\Lambda$CDM sequence;
(2) $n_s$ from 0.85 to 1.25 for the $H_0=70$ $\Lambda$CDM model
($\Omega_{\Lambda}=.66$); (3) $\Omega_B{\rm h}^2$ from 0.003 to 0.05
for the $H_0=70$ $\Lambda$CDM model; (4) $H_0$ from 50 to 65,
$\Omega_{k}$ from 0 to 0.84 for the untilted oCDM sequence; (5) the
same for a $n_s=0.9$ oCDM sequence, clearly at odds with the data; (6)
$H_0$=50 sequence with neutrino fractions varying from 0.1 to 0.95;
(7) shows an isocurvature CDM sequence with positive isocurvature
tilts ranging from 0 to 0.8; (8) shows that sample defect ${\cal
C}_\ell$'s from Pen, Seljak and Turok (1997) do not
fare well compared with the current data; ${\cal C}_\ell$'s from
(Allen \etal 1997) are similar. The bottom right panel is
extended to low values to show the magnitude of secondary fluctuations
from the thermal SZ effect for the $\Lambda$CDM model. The kinematic
SZ ${\cal C}_\ell$ is significantly lower. Dusty emission from early
galaxies may lead to high signals, but the power is concentrated at
higher $\ell$, with possibly a weak tail because galaxies are
correlated extending into the $\ell \lta 2000$ regime. }
\label{fig:CLth}
\end{figure}

The arena in which CMB theory battles observation is the anisotropy
power spectrum in multipole space, as in 
Figs.~\ref{fig:CLdat},\ref{fig:CLth},  which show how primary ${\cal
C}_\ell$'s vary with some of these cosmic parameters. Here ${\cal
C}_\ell \equiv \ell (\ell +1) \avrg{\vert (\Delta T/T)_{\ell m}\vert^2
}/(2\pi )$. The ${\cal C}_\ell$'s are normalized to the 4-year {\it
dmr}(53+90+31)(A+B) data (Bennett \etal 1996a, Bond 1995, Bond \&
Jaffe 1997). The arena for LSS theory battling observations is the
${\cal P}_\Phi$ of Fig.~\ref{fig:probes}. (Usually it is ${\cal
P}_\rho/k^3 \sim k{\cal P}_\Phi$ which is plotted.)

For a given model, the early universe ${\cal P}_{\Phi}$ is uniquely
related to late-time power spectrum measures of relevance for the CMB,
such as the quadrupole ${\cal C}_2^{1/2}$ or averages over
$\ell$-bands B, $\avrg{{\cal C}_\ell}_B^{1/2}$, and to LSS measures,
such as the {\it rms} density fluctuation level on the $8\hmpc$
(cluster) scale, $\sigma_8$, so any of these can be used in place of
the primordial power amplitudes in the parameter set.  In inflation,
the ratio of gravitational wave power to scalar adiabatic power is
${\cal P}_{GW}/{\cal P}_{\Phi}\approx -(100/9)\nu_t/(1-\nu_t/2)$, with
small corrections depending upon $\nu_s-\nu_t$ (Bond 1994, 1996).
If such a relationship is assumed, the parameter count is lowered by
one.

\subsection{Fluctuation Freedom in Inflation}
\label{secinflation} 

 Many variants of the basic inflation theme have been proposed,
sometimes with radically different consequences for ${\cal P}_\Phi (k)
\sim k^{1-n_s(k)}$, and thus for the CMB sky, which is used in fact to
highly constrain the more baroque models.  A rank-ordering of
inflation possibilities: (1) adiabatic curvature fluctuations with
nearly uniform scalar tilt over the observable range, slightly more
power to large scales ($0.8 \lta n_s \lta 1$) than ``scale
invariance'' ($n_s=1$) gives, a predictable nonzero gravity wave
contribution with tilt similar to the scalar one, and tiny mean
curvature ($\Omega_{tot}\approx 1$); (2) same as (1), but with a tiny
gravity wave contribution; (3) same as (1) but with a subdominant
isocurvature component of nearly scale invariant tilt (the case in
which isocurvature dominates is ruled out); (4) radically broken scale
invariance with weak to moderate features (ramps, mountains, valleys)
in the fluctuation spectrum (strong ones are largely ruled out); (5)
radical breaking with non-Gaussian features as well; (6) ``open''
inflation, with quantum tunneling producing a negatively-curved
(hyperbolic) space which inflates, but not so much as to flatten the
mean curvature ($d_c \sim (Ha)^{-1}$, not $\gg (Ha)^{-1}$, where
$d_c\equiv H_0^{-1}\vert \Omega_{k} \vert^{-1/2}$); (7) quantum
creation of compact hyperbolic space from ``nothing'' with volume
$d_T^3$ which inflates, with $d_T \sim (Ha)^{-1}$, not $\gg
(Ha)^{-1}$, and $d_T$ of order $d_c$; (8) flat ($d_c=\infty $)
inflating models which are small tori of scale $d_T$ with $d_T$ a few
$(Ha)^{-1}$ in size. It is quite debatable which of the cases beyond
(2) are more or less plausible, with some claims that (4) is
supersymmetry-inspired, others that (6) is not as improbable as it
sounds. Of course, how likely {\it a priori} the cases (7) and (8) of
most concern to us here is completely unknown, but it is the
theorists' job to push out the boundaries of the inflation idea and
use the data to select what is allowed.

\subsection{LSS Constraints on the Power Spectrum } \label{secLSSpspec}

We have always combined CMB and LSS data in our quest for viable
models. Fig.~\ref{fig:probes} shows how the two are connected. DMR
normalization precisely determines $\sigma_8$ for each model
considered; comparing with the $\sigma_8 \sim 0.6\Omega_{nr}^{-0.56} $
target value derived from cluster abundance observations severely
constrains the cosmological parameters defining the models. In
Fig.~\ref{fig:probes}, this means the COBE-normalized ${\cal
P}_{\Phi}(k)$ must thread the ``eye of the needle'' in the
cluster-band.

Similar constrictions arise from galaxy-galaxy and cluster-cluster
clustering observations: the shape of the linear ${\cal P}_\Phi $ must
match the shape reconstructed from the data. The reconstruction shown
is from Peacock (1997).  The clustering observations are roughly
compatible with an allowed range $0.15 \lta \Gamma + \nu_s/2 \lta
0.3$, where $\Gamma \approx \Omega_{nr} \, {\rm h} \,
[\Omega_{er}/(1.68\Omega_{\gamma})]^{-1/2} \,
e^{-(\Omega_B(1+\Omega_{nr}^{-1}(2{\rm h})^{1/2}) -0.06)}$
characterizes the density transfer function shape. The sCDM model has
$\Gamma \approx 0.5$. To get $\Gamma + \nu_s/2$ in the observed range
one can: lower ${\rm h}$; lower $\Omega_{nr}$ ($\Lambda$CDM, oCDM);
raise $\Omega_{er}$, the density parameter in relativistic particles
($1.68\Omega_{\gamma}$ with 3 species of massless neutrinos and the
photons), \eg as in $\tau$CDM, with a decaying $\nu$ of lifetime
$\tau_d$ and $\Gamma \approx 1.08 \Omega_{nr} {\rm h}(1 + 0.96 (m_\nu
\tau_d /{\rm keV~ yr})^{2/3})^{-1/2}$; raise $\Omega_B$; tilt $\nu_s <
0$ (tCDM), for standard CDM parameters, \eg $0.3 \lta n_s \lta 0.7$
would be required. Adding a hot dark matter component gives a power
spectrum characterized by more than just $\Gamma$. In the post-COBE
era, all of these models that lower $\Gamma + \nu_s/2$ have been under
intense investigation to see which, if any, survive as the data
improve.

\subsection{Cosmological Radiative Transport} \label{sectransport}

Cosmological radiative transfer is on a firm theoretical
footing. Together with a gravity theory (invariably Einstein's general
relativity, but the CMB will eventually be used as a test of the
gravity theory) and the transport theory for the other fields and
particles present (baryons, hot, warm and cold dark matter, coherent
fields, \ie ``dynamical'' cosmological ``constants'', {\it etc.}), we
propagate initial fluctuations from the early universe through photon
decoupling into the (very) weakly nonlinear phase, and predict {\it
primary anisotropies}, those calculated using either linear
perturbation theory (\eg for inflation-generated fluctuations), or, in
the case of defects, linear response theory. The sources driving their
development are all proportional to the gravitational potential
$\Phi$: the ``naive'' Sachs-Wolfe effect, $\Phi /3$; photon bunching
and rarefaction (acoustic oscillations), ${1\over 4} {\delta
\rho_\gamma \over \rho_\gamma }$, responsible for the adiabatic
${1\over 3} {\delta \rho_B \over \rho_B }$ effect and the isocurvature
effect; linear-order Thompson scattering (Doppler), $\sigma_T
\bar{n}_e {\bf v}_e \cdot \hq$, with $\sigma_T$ the Thomson cross section,
${\bf v}_e$ and $\bar{n}_e$ the electron velocity and density, and $\hq$
the photon direction; the (line-of-sight) integrated Sachs-Wolfe
effect, $\sim 2\int_{l.o.s.}  \dot{\Phi}$; there are also subdominant
anisotropic stress and polarization terms. For primary tensor
anisotropies, the sources are the two polarization states of gravity
waves, $\half \dot{h}_{+, \times}$; again there are subdominant
polarization terms.

Spurred on by the promise of percent-level precision in cosmic
parameters from CMB satellites (\S~\ref{seccmbfuture}), a considerable
fraction of the CMB theoretical community with Boltzmann transport
codes compared their approaches and validated the results to ensure
percent-level accuracy up to $\ell \sim 3000$ (COMBA 1995).  An
important goal for COMBA was speed, since the parameter space we wish
to constrain has many dimensions.  Most groups have solved
cosmological radiative transport by evolving a hierarchy of coupled
moment equations, one for each $\ell$. Although the equations and
techniques were in place prior to the COBE discovery for scalar modes,
and shortly after for tensor modes, to get the high accuracy with
speed has been somewhat of a challenge. There are alternatives to the
moment hierarchy for the transport of photons and neutrinos. In
particular the entire problem of photon transport reduces to integral
equations in which the multipoles with $\ell >2$ are expressed as
history-integrals of metric variables, photon-bunching, Doppler and
polarization sources. The fastest COMBA-validated code, ``CMBfast'',
uses this method (Seljak \& Zaldarriaga 1996), is publicly available
and widely used (\eg to generate some of the power spectra in
Fig.~\ref{fig:CLth}).

\subsection{Secondary Anisotropies} \label{secsec}

Although hydrodynamic and radiative processes are expected to play
important roles around collapsed objects and may bias the galaxy
distribution relative to the mass ({\it gastrophysics} regime in
Fig.~\ref{fig:probes}), a global role in obscuring the early universe
fluctuations by late time generation on large scales now seems
unlikely. Not too long ago it seemed perfectly reasonable based on
extrapolation from the physics of the interstellar medium to the
pregalactic and intergalactic medium to suppose hydrodynamical
amplification of seed cosmic structure could create the observed
Universe. The strong limits on Compton cooling from the COBE FIRAS
experiment (Fixsen \etal 1997), in energy ${\delta E_{Compton\ cool} /
E_{cmb}} = 4y < 6.0 \times 10^{-5}$ (95\% CL), constrain the product
$f_{exp}R_{exp}^2$ of filling factor $f_{exp}$ and bubble formation
scale $R_{exp}$, to values too small for a purely hydrodynamic
origin. If supernovae were responsible for the blasts, the
accompanying presupernova light radiated would have been much in
excess of the explosive energy (more than a hundred-fold), leading to
much stronger restrictions (\eg Bond 1996).

Nonetheless significant ``secondary anisotropies'' are expected. These
include: linear weak lensing, dependent on the 2D tidal tensor, a
projection of the 3D tidal tensor $\partial^2 \Phi /\partial
x^i\partial x^j$; the Rees-Sciama effect, $2\int_{l.o.s.}
\dot{\Phi}_{NL}$, dependent upon the gravitational potential changes
associated with nonlinear structure formation; nonlinear Thompson
scattering, $\sigma_T \delta {n}_e {\bf v}_e \cdot \hq$, dependent upon
the fluctuation in the electron density $\delta {n}_e$ as well as
${\bf v}_e$, and responsible for the quadratic-order (Vishniac) effect and
the ``kinematic'' Sunyaev-Zeldovich (SZ) effect (moving cluster/galaxy
effect); the thermal SZ effect, associated with Compton cooling,
$\int_{l.o.s.}  \psi_K (x ) \delta (n_e T_e) $, where $\psi_K (x )$ is
a function of $x=E_\gamma/T_\gamma$ passing from $-2$ on the
Rayleigh-Jeans end to $x$ on the Wein end, with a null at $x=2.83$
(\ie $1863 \mu {\rm m}$ or 161 GHz); pregalactic or galactic dust
emission, $ \sim \int_{l.o.s.}  \psi_{dust}(x_d)\rho_d$, dependent
upon the distribution of the dust density $\rho_d$ and temperature $
T_d$ through a function $\psi_{dust}$ of $x_d=E_\gamma/T_d$.

Secondary anisotropies may be considered as a nuisance foreground to
be subtracted to get at the primary ones, but they are also invaluable
probes of shorter-distance aspects of structure formation theories,
full of important cosmological information. The $k$-space range they
probe is shown in Fig.~\ref{fig:probes}.  The effect of lensing is to
smooth slightly the Doppler peaks and troughs of
Fig.~\ref{fig:CLth}. ${\cal C}_\ell$'s from quadratic nonlinearities
in the gas at high redshift are concentrated at high $\ell$, but for
most viable models are expected to be a small contaminant. Thomson
scattering from gas in moving clusters also has a small effect on
${\cal C}_\ell$ (although it should be measurable in individual
clusters).  Power spectra for the thermal SZ effect from clusters are
larger (Bond \& Myers 1996); the example in the bottom panel of
Fig.~\ref{fig:CLth} is for an untilted $H_0=70$ COBE-normalized
$\Lambda$CDM model, with $\bar{y}\sim 2\times 10^{-6} \, (\Omega_B{\rm
h}^2/0.025)$, still small {\it c.f.} the FIRAS constraint. (Here and
in the following, when $H_0$ values are given, the units $\kms
\mpc^{-1}$ are implicit.) Although ${\cal C}^{(SZ)}_\ell$ may be
small, because the power for such non-Gaussian sources is concentrated
in hot or cold spots the signal is detectable, in fact has been for
two dozen clusters now at the $>5$ sigma level, and indeed the SZ
effect will soon be usable for cluster-finding.  ${\cal C}_\ell$ for a
typical dusty primeval galaxy model is concentrated at higher $\ell$
associated with galaxy sizes, although a small contribution associated
with clustering extends into the lower $\ell$ range. These dusty
anisotropies are now observable with instrumentation on submm
telescopes (\eg SCUBA on the James Clerk Maxwell Telescope on Mauna
Kea, with the $k$-space filter shown in Fig.~\ref{fig:probes}).

\section{CMB Parameter Estimation, Current and Future}

\subsection{Comparing and Combining CMB Experiments}\label{seccombine} 

We have progressed from the tens of pixels of early $\Delta T/T$
experiments through thousands for DMR (Bennett  \etal  1996a) and SK95
(Netterfield \etal 1996), soon tens of thousands for long duration balloon
experiments (LDBs) and eventually millions for the MAP (Bennett \etal 1996b)
and Planck (Bersanelli  \etal 1996) satellites. Finding nearly optimal
strategies for data projection, compression and analysis which will
allow us to disentangle the primary anisotropies from the Galactic and
extragalactic foregrounds and from the secondary anisotropies induced
by nonlinear effects will be the key to realizing the
theoretically-possible precision on cosmic parameters and so to
determine the winners and losers in theory space.  Particularly
powerful is to combine results from different CMB experiments and
combine these with LSS and other observations.  Application of the
same techniques to demonstrate self-consistency and cross-consistency
of experimental results is essential for validating conclusions drawn
from the end-product of data analysis, {\it e.g.}, the power spectra
in bands as shown in Fig.~\ref{fig:CLdat} and the cosmic
parameters they imply.

Current band-powers are shown in the upper panel of
Fig.~\ref{fig:CLdat}. The first lesson of
Figs.~\ref{fig:CLdat},\ref{fig:CLth} is that, in broad brush stroke,
smaller angle CMB data (\eg SP94, SK95, MSAM, MAX) are consistent with
COBE-normalized ${\cal C}_\ell$'s for the untilted inflation-based
models. It is possible that some of the results may still include
residual contamination, but it is encouraging that completely different
experiments with differing frequency coverage are highly correlated
and give similar bandpowers, {\it e.g.} DMR and FIRS (\eg Bond 1996), SK95
and MSAM (Netterfield \etal 1996, Knox \etal 1998). Lower panels
compress the information into 9 optimal bandpower estimates derived
from all of the current data (see Bond, Jaffe \& Knox 1998b for
techniques).

The few data points below $\ell \lta 20$ are mainly from COBE's DMR
experiment. Clearly the $\ell$-range spanned by DMR is not large
enough to fix well the cosmological parameter variations shown in the
right panels, but combining CMB anisotropy experiments probing
different ranges in $\ell$-space improves parameter estimates
enormously because of the much extended baseline: it is evident that
$n_s$ can be reasonably well determined, low $\Omega$ open models
violate the data, but $\Omega_\Lambda$ cannot be well determined by
the CMB alone.

\subsection{DMR and Constraints on Ultra-large Scale Structure}\label{secULSScurrent} 

DMR is fundamental to analyses of the VLSS region and ULSS region, and
is the data set that is the most robust at the current time. The
average noise in the 53+90+31 GHz map is about 20$\mu K$ per {\it
fwhm} beam ($\sim 7^\circ$), and there are about 700 of these
resolution elements outside of the Galactic disk cut (about 4000
2.6$^\circ$ DMR pixels with 60$\mu K$ noise). The signal is about
37$\mu K$ per beam: \ie there is a healthy signal-to-noise ratio. The
signal-to-noise for widespread modes (\eg multipoles with $\ell \lta
15$) is even better. Indeed, even with the much higher precision MAP
and Planck experiments we do not expect to improve the results on the
COBE angular scales greatly because the 4-year COBE data has
sufficiently large signal-to-noise that one is almost in the cosmic
variance error limit (due to realization to realization fluctuations
of skies in the theories) which cannot be improved upon no matter how
low the experimental noise.

\begin{figure}

\begin{center}
\epsfysize=5.5in\leavevmode\epsffile{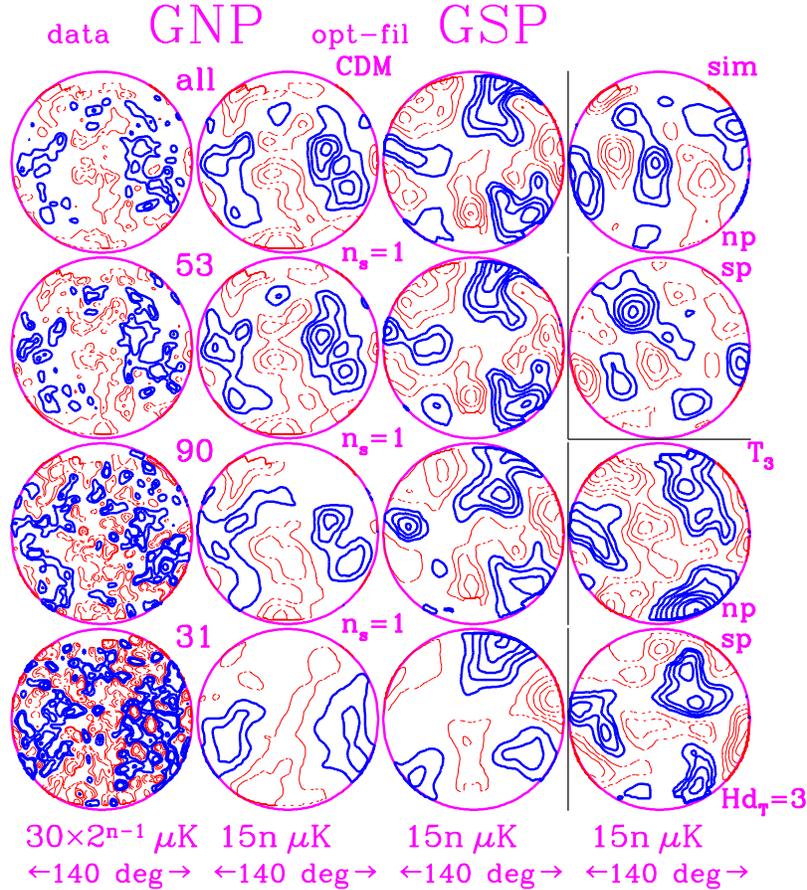}
\end{center}
\vspace{-0.4in}
\caption{\figcapsize The first column shows unfiltered $140^\circ$
diameter {\it dmr} $A$+$B$ maps centered on the North Galactic Pole,
the second shows them after Wiener-filtering (with monopole, dipole
and quadrupole removed), the third the South Pole version, with the
$nth$ contour as noted and negative contours heavier than positive
ones.  The Wiener maps use a model which fits the correlation function
and amplitude of the DMR data (specifically, the $n_s$=1 sCDM model
was used, but insensitive to even rather significant variations). The
maps have been smoothed by a $1.7^\circ$ Gaussian filter.  {\it all}
is 53+90+31$A$+$B$. Although higher noise results in filtering on
greater angular scales, the large scale features of all maps are the
same. This is also borne out by detailed statistical comparisons map
to map. The last column shows some theoretical realizations, after
optimal filtering. The first two rows are the NGP and SGP for a
$n_s$=1 CDM model. The lower two rows are for a 3-torus topology, with
repetition length $d_T=9000 \hmpc $, 1.5 times the horizon radius, in
all three directions, a model strongly ruled out because of the high
degree of positive correlation between the North and Southern
hemispheres that the periodicity induces (Bond, Pogosyan \& Souradeep
1998). Highly correlated patterns also exist for small compact
hyperbolic models and lead to constraints on manifold size.}
\label{fig:optfilNP4}
\end{figure}

Wiener-filtered maps shown in Fig.~\ref{fig:optfilNP4} give the
statistically-averaged signal given the data and a best-fit signal
model. These optimally-filtered maps are insensitive to modest
variations in the assumed theory.  The robustness of features in the
maps as a function of frequency and the weak frequency dependence in
the bandpowers are strong arguments that what is observed is on the
sky with a primary anisotropy origin, made stronger by the compatible
amplitudes and positive cross-correlations with the FIRS and Tenerife
data sets.

Recall the ``beyond our horizon'' land in Fig.~\ref{fig:probes} is
actually partly accessible because long waves contribute gentle
gradients to our observables. The DMR data is well suited to probe
this regime.  Constraints on such ``global parameters'' as average
curvature from COBE are not very good. Obviously it is much preferred
to use the smaller angle data on the acoustic peak positions.  The
COBE data can be used to test whether radical broken scale invariance
results in a huge excess or deficit of power in the COBE $k$-space
band, {\it e.g.}, just beyond $k^{-1} \sim H_0^{-1}$, but this has not
been much explored. The remarkable non-Gaussian structure predicted by
stochastic inflation theory would likely be too far beyond our horizon
for its influence to be felt. The bubble boundary in hyperbolic
inflation models may be closer and its influence shortly after quantum
tunneling occurred could possibly have observable consequences for the
CMB. Theorists have also constrained the scale of topology in simple
models (Fig.~\ref{fig:optfilNP4}). Bond, Pogosyan \& Souradeep (1997,
1998) find the torus scale is $d_T/2 > 1.1 (2H_0^{-1}) = 6600 \hmpc$
from DMR for flat equal-sided 3-tori at the $95\%$ confidence limit,
slightly better than other groups find since full map statistics were
used.  The constraint is not as strong if the repetition directions
are asymmetric, $> 0.7 (2 H_0^{-1})$ for 1-tori from DMR. It is also
not as strong if more general topologies are considered, \eg the large
class of compact hyperbolic topologies (Bond, Pogosyan \& Souradeep
1997, 1998, Cornish \etal 1996, 1998, Levin \etal 1997, 1998).

\subsection{Cosmic Parameters from All Current CMB Data}\label{seccmbcurrent} 

We have undertaken full Bayesian statistical analysis of the 4 year
DMR (Bennet \etal 1996a), SK94-95 (Netterfield \etal 1996) and SP94
(Gundersen \etal 1995) data sets, taking into account all correlations
among pixels in the data and theory (Bond \& Jaffe 1997). Other
experiments available up to March 1998 were included by using their
bandpowers as independent points with the Gaussian errors shown in
Fig.~\ref{fig:CLdat}. We have shown this approximate method works
reasonably well by comparing results derived for DMR+SP94+SK95 with
the full analysis with those using just their bandpowers (Jaffe, Knox
\& Bond 1997). We have also shown that a significant improvement in
accuracy is possible if instead of the average and 1 sigma limits on
the experimental bandpowers ${\cal C}_B$ or on ${\cal C}_B^{1/2}$, one
uses $\ln ({\cal C}_B + x_B)$, where $x_B$ is related to the noise of
the experiment (Bond, Jaffe \& Knox 1998b). This includes some of the
major non-Gaussian deviations in the bandpower likelihood functions;
results using this more accurate approach, and incorporating the very
recent CAT98 and QMAP data, will be reported elsewhere (Bond, Jaffe \&
Knox 1998c, in preparation). Other groups have also calculated
parameter constraints using the ${\cal C}_B^{1/2}$ bandpower approach
(\eg Lineweaver \& Barbosa 1997, Hancock \& Rocha 1997, Lineweaver
1998).

With current errors on the data, simultaneously exploring the entire
parameter space of \S~\ref{seccosmicparam} is not useful, so we
restricted our attention to various subregions of $\{ \Omega_B{\rm
h}^2 ,\Omega_{cdm}{\rm h}^2,\Omega_{hdm}{\rm h}^2, \Omega_{k}{\rm
h}^2, \Omega_{\Lambda}{\rm h}^2, \nu_s, \nu_t, \sigma_8 \}$, such as
$\{ \sigma_8 , n_s, {\rm h}\, | \, {\rm fixed} \, t_0, \Omega_B {\rm
h}^2 \}$, where $\Omega_{k}$=0 and $\Omega_{\Lambda}$ is a function of
${\rm h}t_0$ or $\Omega_{\Lambda}$=0 and $\Omega_{k}>0$ is a function
of ${\rm h}t_0$.  The age of the Universe, $t_0$, was chosen to be 11,
13 or 15~Gyrs. A recent estimate for globular cluster ages with the
Hipparcos correction is $11.5\pm 1.3~{\rm Gyr}$ (Chaboyer \etal 1998),
with perhaps another Gyr to be added associated with the delay in
globular cluster formation, so 13 Gyr is a good example. We considered
the ranges $0.5\le n_s \le 1.5$, $0.43 \le {\rm h}\le 1$, and $0.003
\le \Omega_B {\rm h}^2\le 0.05$. The old ``standard'' nucleosynthesis
estimate was $\Omega_B {\rm h}^2=0.0125$, but the preferred one is now
$0.025$. We assumed reheating occurred sufficiently late to have a
negligible effect on ${\cal C}_\ell$, although this is by no means
clear.  ${\cal C}_\ell$'s for sample restricted parameter sequences
are shown in Fig.~\ref{fig:CLth}.  We made use of signal-to-noise
compression of the data (by factors of 3) in order to make the
calculations of likelihood functions such as ${\cal L}(\sigma_8 , n_s,
{\rm h}\, | \, {\rm fixed} \, t_0, \Omega_B {\rm h}^2)$ more tractable
(without loss of information or accuracy).

The $n_s$ constraints are quite good.  If $\sigma_8$ is marginalized
for the tilted $\Lambda$CDM sequence with $H_0$=50, with DMR only the
primordial index is $n_s$ = $1.02^{+.23}_{-.25}$ with no gravity waves
and $\nu_t$=0, and $1.02^{+.23}_{-.18}$ with gravity waves and
$\nu_t$=$\nu_s$, rather encouraging for the nearly scale invariant
models preferred by inflation theory.  Because the gravitational
potential changes at late time with $\Lambda \ne 0$, the integrated
Sachs-Wolfe effect gives more power in ${\cal C}_\ell$ at small
$\ell$, so the preferred $n_s$ steepens to compensate. When $\Lambda$
is marginalized in the 13 Gyr tilted $\Lambda$CDM sequence, $n_s=1.17
\pm 0.31$ is obtained. For this sequence, when all of the current CMB
data are used we get $1.02^{+.05}_{-.03}$ for $H_0=50$ (and
$\Omega_{\Lambda}=0$, the tilted sCDM model sequence) and
$1.00^{+.04}_{-.04}$ for $H_0=70$ (and
$\Omega_{\Lambda}=0.66$). Marginalizing over $H_0$ (\ie $\Lambda$)
gives $1.01^{+.05}_{-.04}$ with gravity waves included,
$0.98^{+.08}_{-.06}$ if they are not. The marginalized 13 Gyr tilted
oCDM sequence gives $1.00^{+.05}_{-.05}$.

\begin{figure}[htbp]
\begin{center}
\epsfysize=6.0in\leavevmode\epsffile{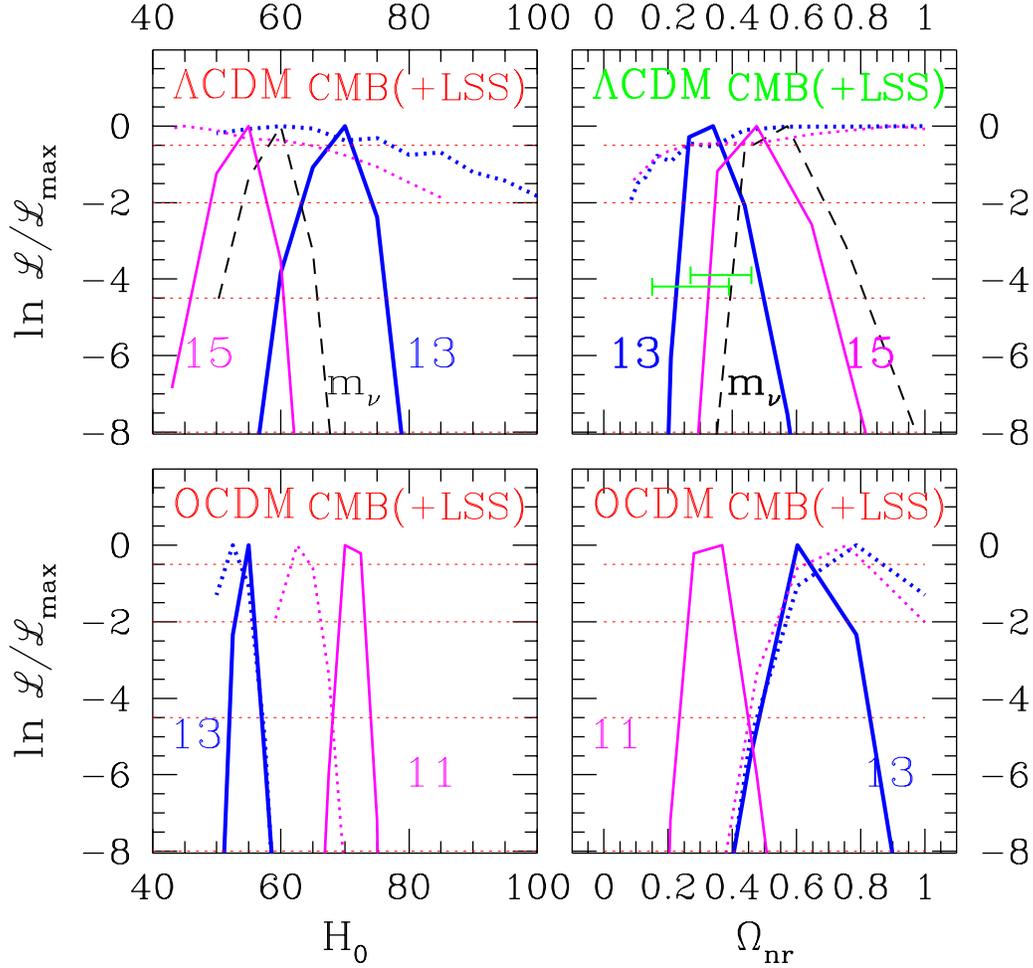}
\end{center}
\vspace{-0.3in}
\caption{\figcapsize Likelihood curves for fixed-age $\Lambda$CDM and
oCDM sequences, marginalized over $\sigma_8$ and $n_s$. A Gaussian
approximation to the likelihood places 1,2,3$\sigma$ at the horizontal
dashed lines.  The 1$\sigma$ ranges are explicitly given in the
text. The dotted curves are for CMB only, solid for CMB+LSS. The right
panels are equivalent to the left, but translated to $\Omega_{nr}$
($1-\Omega_\Lambda$ for $\Lambda$CDM, $\Omega_{tot}$ for oCDM). The
curves shown are for no GW, but there is little difference if GW are
included.  The sequence labelled $m_\nu$ is for the 13 Gyr $\Lambda$hCDM sequence
with a fixed $\Omega_{m\nu}/\Omega_{nr}$ ratio of 0.2, and 2 degenerate
neutrino species. Even this case favours slightly a nonzero
$\Omega_\Lambda$. In the upper left panels, the 68\% confidence limits
for the values of $\Omega_m$ estimated using the luminosity-distance
relation for Type I supernovae from Perlmutter \etal and Craig \etal
are also shown. Note that the CMB data alone slightly favours a value
of $\Omega_{tot}<1$.  The absolute likelihood for the CMB+LSS data
strongly favours the $\Lambda$CDM over the oCDM sequences. When just
the fully analyzed DMR+SK95+SP94 data are used with LSS, the
marginalized results for $\Lambda$CDM are remarkably similar, with the
long baseline between DMR and SK95 fixing the freedom in $n_s$
(although values of $n_s$ about 1.1 are now preferred). When only DMR
is used along with LSS, $n_s$ is not nailed down, and the resulting
freedom implies $\Omega_{nr}=1$ models are not disfavoured. The
horizontal error bars in the upper right panel show the $1\sigma$
range of $\Omega_{nr}$ for $\Omega_{nr}+\Omega_\Lambda=1$ models
inferred from the supernova Ia observations of 
Perlmutter \etal 1998 (upper) and Reiss \etal 1998 (lower).}
\label{fig:like}
\end{figure}

$H_0$ and $\Omega_\Lambda$ for fixed age are not that well determined
by the CMB data alone, as can be seen from the dotted lines in
Fig.~\ref{fig:like}. After marginalizing over all $n_s$, we get $H_0 <
75$ at $1\sigma$, but effectively no constraint at 2$\sigma$. The
strong dependence of the position of the acoustic peaks on $\Omega_k$
means that the oCDM sequence is better restricted: $\Omega_{tot}\sim
.7$ is preferred; for the 13 Gyr sequence this gives $H_0 \approx 53$
and for the 11 Gyr sequence $H_0 \approx 65$. 

Calculations of defect models ({\it e.g.} strings and textures) give
${\cal C}_\ell$'s that do not have the prominent peak that the data seem
to indicate (Pen, Seljak \& Turok 1997, Allen \etal 1997).

\subsection{Cosmic Parameters from Current LSS plus CMB Data}\label{seccmbLSScurrent} 

Combining LSS and CMB data gives more powerful discrimination among
the theories, as Fig.~\ref{fig:probes} illustrates visually and
Fig.~\ref{fig:like} shows quantitatively. The approach we use here and
in Bond \& Jaffe (1997) to add LSS information to the CMB
likelihood functions is to design prior probabilities for $\Gamma
+\nu_s /2$ and $\sigma_8 \Omega_{nr}^{0.56} $, reflecting the current
observations, but with flexible and generous non-Gaussian and
asymmetric forms to ensure the priors can encompass possible
systematic problems in the LSS data. For example, our choice for
$\sigma_8 \Omega_{nr}^{0.56} $ was relatively flat over the 0.45 to
0.65 range. (Explicitly we used $0.55^{+.02+.15}_{-.02-.08}$, with the
two error bars giving a Gaussian and a top hat error so that the net
result is generously flat over the total $\pm 1\sigma$ range. For
$\Gamma +\nu_s /2$, we used $0.22^{+.07+.08}_{-.04-.07}$. Using the
Peacock 1997 reconstructed linear power spectrum shown in
Fig.~\ref{fig:probes} would give more stringent constraints for the shape 
\eg Gawiser \& Silk 1998.)

Using all of the current CMB data and the LSS priors, for the 13 Gyr
$\Lambda$CDM sequence with gravity waves included, we get
$n_s=1.00^{+.05}_{-.03}$ and $H_0=72\pm 3$ ($\Omega_{\Lambda}\approx
0.7$), respectively, when $H_0$ and $n_s$ are marginalized; with no
gravity waves, $0.96^{+.07}_{-.05}$ and $H_0=70\pm 3$ are obtained;
and for an $\Lambda$hCDM sequence, with a fixed ratio
$\Omega_{hdm}/\Omega_{nr}=0.2$ for two degenerate massive neutrino
species, $n_s\approx 0.97^{+.02}_{-.02}$ and $H_0 \approx
57^{+5}_{-3}$ are obtained, revealing a slight preference for
$\Omega_{\Lambda} \sim 0.3$.  For the 15 Gyr $\Lambda$CDM sequence,
the tilts remain nearly scale invariant and $\Omega_{\Lambda}$ near
0.6: $0.98^{+.04}_{-.03}$ and $H_0=57 \pm 3$ ($\Omega_{\Lambda}\approx
0.6$) with gravity waves, $0.95^{+.05}_{-.05}$ and $54 \pm 3$
($\Omega_{\Lambda}\approx 0.5$) without. 

For the 13 Gyr oCDM sequence, the likelihood peak for the CMB+LSS
data is shifted relative to using the CMB data alone because the best
fit CMB-only models have $\sigma_8$ too low compared with the cluster
abundance requirements. Although the $H_0\approx 54^{+1}_{-1}$ value
($\Omega_{tot} \approx 0.6$) is close to the CMB-only one, the maximum
likelihood is significantly below the $\Lambda$CDM one. $H_0$ is
larger for the 11 Gyr oCDM sequence, but $\Omega_{tot}$ is about the
same, and the likelihood is still low.

Should these small error bars be taken seriously?  It seems unlikely
that $\sigma_8$ from cluster abundances will change much; and, as we
have seen, the DMR results are quite robust.  Although largely driven
by just the DMR plus LSS results, the smaller angle CMB results lock
in the tilt, and as the CMB data improves some adjustment might occur,
but not a drastic one unless we have made a major misinterpretation in
the nature of the CMB signals observed at intermediate angles.

If we were to marginalize over $t_0$ as well, it is clear that $H_0$
would not be as well determined, but $n_s$ and either $\Omega_k$ or
$\Omega_\Lambda$ would be. If the parameter space is made even larger,
near degeneracies among some cosmic parameters become important for
CMB data alone, and these are only partially lifted by the LSS data
(\eg Efstathiou \& Bond 1998). In particular, this restricts the
ultimate accuracy that can be achieved in the simultaneous
determination of $\Omega_k$ and $\Omega_\Lambda$. This will become an
issue when the quality of the CMB data improves, as described in the
next subsection, but for now one must bear in mind the constrained
space used when interpreting the current precision quoted on parameter
estimation.

\subsection{Cosmic Parameters from the CMB Future} \label{seccmbfuture}

The expected error bars on the power spectrum from MAP and Planck
(Bond, Jaffe \& Knox 1998a, Bond, Efstathiou \& Tegmark 1997,
hereafter BET) shown in Fig.~\ref{fig:CLdat} illustrate that even quite
small differences in the theoretical ${\cal C}_\ell$'s and thus the
parameters can be distinguished. Quite an industry has developed
forecasting how well future balloon experiments (Maxima, Boomerang,
ACE, Beast, Top Hat), interferometers (VSA, CBI, VCA) and especially
the satellites MAP and Planck could do in measuring the radiation
power spectrum and cosmological parameters if foreground contamination
is ignored (Knox 1995, Jungman \etal 1996, BET, Zaldarriaga, Spergel,
\& Seljak 1997, White, Carlstrom \& Dragovan 1997).  Forecasts like
these were quite influential in making the case for MAP and Planck.

\begin{table}
  \begin{center}
\begin{tabular}{|c|c|c|cc|cc|ccc|ccc|}
\hline
 & Max & TH & BstI & II &  Bm&  Bm & &MAP & & &Pl & \\
$f_{sky}$ & .01 & .028 & .067 & .067 & .02 & .02 & .67  & .67  & .67 & .67  & .67  & .67  \\
$\ell_{cut}$  & 20 & 12 & 6 &6 & 12 & 12 & 2 & 2 & 2 & 2 & 2 & 2  \\
$\nu$ & all & all & 40 & 90 & 90 & 150 & 90 &60 & 40 &100 & 150 &
220   \\
$\theta_{fwhm}$ & 12 & 20 & 19 & 9  & 20 & 12 & 13  &18 &32 & 14.5 & 10 & 6.6  \\  
$\sigma_{Npix}$ & 24 & 18.4 & 31  & 100 & 21 & 35 & 34 & 25 & 14 & 3.4 & 3.6 & 3.2 \\
\hline
\end{tabular}
\caption{\figcapsize The noise is per {\it fwhm} pixel, in $\mu$K. Max
is Maxima, TH is TopHat, and Bst is Beast, which also has a 30 GHz
channel ($26^\prime$, 53$\mu$K) included in the analysis. Bm is
Boomerang, which also has 220 and 430 GHz channels not included in the
analysis. The North America test flights covered .008 and .003 of the
sky, and had {\it fwhm} of $35^\prime$ and $ 19^\prime$,
respectively. Boost is a bolometer based version of Beast that is an
example of an ultra-long duration balloon experiment, of order 100
days, with channels at 100 GHz ($9^\prime$, 17$\mu$K), 140
GHz($6^\prime$, 24$\mu$K), ($4^\prime$, 37$\mu$K), covering .07 of the
sky, with $\ell_{cut}\approx 6$. The Planck numbers are for the HFI
channels; a 350 GHz ($4^\prime$, 43$\mu$K) was also included in the
analysis. See BET for the 3 channels of LFI used. }
\end{center}
\label{tab:exptparams} 

  \begin{center}
\begin{tabular}{|c|c|c|c|cc|}
\hline
Param &  Current &MaxTHBoom & MAP & Planck & Planck  \\
 &  range &+BeastI+dmr&  &LFI  & HFI\\
\hline
$f_{sky}$ & & 0.07 & 0.67 & 0.67 & 0.67  \\
\hline
 $\delta n_s$   &(0.5--1.5) & .07 &    .04 & .01 &.006  \\
$\delta r_{ts}$ &  (0--1)  & .55      & .24  & .13  &.09 \\
 $\delta \Omega_b{\rm h}^2/\Omega_b{\rm h_0}^2$ & (0.01--0.03) & .11&   .05 &   .016 &.006   \\
 $\delta \Omega_{m}{\rm h}^2/{\rm h_0}^2$& (.2-1) & .20 &  .10  & .04    &  .02  \\
 $\delta \Omega_{\Lambda}{\rm h}^2/{\rm h_0}^2$& (0--0.8) & .46  & .28  & .14  &  .05  \\
 $\delta \Omega_{hdm}{\rm h}^2/{\rm h_0}^2$ &(0--0.3) &.14&    .05&    .04 & .02  \\
  $\tau_C$  &(0.01--1)  &  .26 &   .19 &   .18 & .16 \\
\hline
 $\delta {\rm h}/{\rm h}$ & (40--80)   & .15 &    .11  & .06 &.02 \\
\hline
$\delta \Omega_{k}{\rm h}^2/{\rm h_0}^2$ & (0.2--1.5) & .06 & .04  &    .02  & .007  \\
\hline
\multicolumn{6}{|c|}{Orthogonal Parameter Combinations within $\eps$} \\
\hline
$\eps <0.01$ & 0/9 & 2/9 & 3/9 &  3/9 & 5/9  \\
$\eps <0.1$ & 1/9 & 6/9 &  6/9 &  6/9 & 7/9   \\
\hline
\end{tabular}
\caption{\figcapsize Sample idealized MAP and Planck parameter error
forecasts, for a 9 parameter inflation family of models, with standard
CDM the ``Target Model''. See BET for methods. $\Omega_{\Lambda}{\rm
h}^2$ is determined with $\Omega_k{\rm h}^2$ fixed, and $\Omega_k{\rm
h}^2$ is determined with $\Omega_{\Lambda}{\rm h}^2$ fixed, because of
the angle-distance near-degeneracy (\eg Efstathiou \& Bond 1998); the
other parameters are insensitive to fixing either, or neither. The
ranges for $H_0$, $\Omega_B{\rm h}^2$ are absolute, but the errors are
relative ones. The forecasted errors obviously represent a great leap
forward from current errors and from what is conceivable with non-CMB
probes. Amplitude parameters are highly correlated with $\tau_C$, but
this can be partly broken when other information is included, \eg on
the abundance of clusters. The third column is an optimistic forecast
of what one can do with balloons by combining MAXIMA, TopHat,
Boomerang and BeastI with DMR (see Fig.~\ref{fig:CLdat}). TopHat,
Boomerang, and BeastI would be long duration balloon flights, lasting
about a week over the Antarctic. The parameters used are given in
Table~\ref{tab:exptparams}.  It is unclear that systematics will be
sufficiently small for the LDB experiments to fulfill this promise. }
\label{tab:params} 
  \end{center}
\end{table}

Table~\ref{tab:params} gives some examples of what can be obtained
using only CMB data (BET). The experimental parameters chosen are
given in Table~\ref{tab:exptparams}. The durations chosen were
appropriate for the types of experiments, {\it e.g.}, about a week for
long duration balloon experiments and about two years for satellite
experiments.  The temperature anisotropies were assumed to be
Gaussian-distributed, and among the $> 17$ parameters of
\S~\ref{seccosmicparam}, a restricted 9 parameter space was used: 5
densities, $\{ \Omega_B, \Omega_{nr}, \Omega_{hdm}, \Omega_k ,
\Omega_\Lambda \}{\rm h}^2$, the Compton depth $\tau_C$, the scalar
tilt, $n_s$, the total bandpower for the experiment $\avrg{{\cal
C}_\ell}_B$ in place of ${\cal P}_\Phi (k_n)$, and the ratio of tensor
to scalar quadrupole powers, ${r}_{ts} \equiv {{\cal C}^{(T)}_2 /
{\cal C}^{(S)}_2}$, in place of $\nu_t$. Just like ${\cal
P}_{GW}/{\cal P}_\Phi$, ${r}_{ts}$ is a sensitive function of $\nu_t$,
but also depends on $\nu_s -\nu_t$, $\Omega_{\Lambda}$, \etc (Bond
1996).  In this space, recall that ${\rm h}^2 = \sum_j (\Omega_j{\rm
h}^2 )$ is a dependent quantity.

Except for the integrated Sachs-Wolfe effect at low $\ell$, the
angular pattern of CMB anisotropies now is a direct map of the
projected spatial pattern at redshift $\sim 100$, dependent upon the
cosmological angle-distance relation, which is constant along a line
relating $\Omega_k{\rm h}^2 $ and $ \Omega_\Lambda{\rm h}^2 $ for
fixed $\Omega_{nr}{\rm h}^2$. This defines a near-degeneracy between
$\Omega_k$ and $\Omega_{\Lambda}$ broken only at low $\ell$ where the
large cosmic variance precludes accurate determination of both
parameters simultaneously ({\it e.g.}  BET, Zaldarriaga, Spergel,\&
Seljak 1997, Efstathiou \& Bond 1998, Eisenstein, Hu \& Tegmark
1998). Other cosmological observables are needed to break this
degeneracy. A good example is Type I supernovae. If they are assumed
to be ``standard candles'', then their degeneracy is along lines of
equal luminosity-distance, which is sufficiently different from the
equal angle-distance lines to allow good separate determination.

If the polarization power spectrum can be measured with reasonable
accuracy, errors on some parameter such as $r_{ts}$ would improve
(Zaldarriaga \etal 1997). However the polarization power spectrum is about a
hundred times lower than the total anisotropy, and the gravity wave
induced polarization is substantially tinier than this at the low
$\ell$ needed for $r_{ts}$ improvement. We do not know if the
foreground polarization will hopelessly swamp this signal.

Error forecasts do depend upon the correct underlying theory.  In
Table~\ref{tab:params}, untilted sCDM was chosen as the target model,
but the values shown are indicative of what is obtained with other
targets (BET).  The third column gives errors forecasted for balloon
experiments, the bolometer-based TopHat, Boomerang, and MAXIMA and the
HEMT-based BEAST. (URLs to home pages are given in the references.)
$\ell$-cuts were included to reflect the limited sky coverage these
experiments will have. Adding DMR to extend the $\ell$-baseline
diminishes the forecasted errors.

We adopt the current beam sizes and sensitivities for MAP
and Planck used in BET, improvements over
the original proposal values.  Of the 5 HEMT channels for MAP, BET 
assumed the 3 highest frequency channels, at 40, 60 and 90 GHz, will be
dominated by the primary cosmological signal (with 30 and 22 GHz
channels partly contaminated by bremsstrahlung and synchrotron
emission). MAP also assumes 2 years of observing. For Planck, BET used 
14 months of observing, the 100, 65, 44 GHz channels for the
HEMT-based LFI (but not the 30 GHz channel), and the 100, 150, 220 and
350 GHz channels for the bolometer-based HFI (but not the
dust-monitoring 550 and 850 GHz channels). The highest resolution for
MAP is $13^\prime$ {\it fwhm}, the highest for Planck is $4^\prime$.

These idealized error forecasts do not take into account the cost of
separating the many components expected in the data, in particular
Galactic and extragalactic foregrounds, but there is currently
optimism that the Galactic foregrounds at least may not be a severe
problem \eg (Bersanelli \etal 1996), although low frequency emission
near 100 GHz by small spinning dust grains (Leitch \etal 1997, Draine
\& Lazarian 1998) may emerge as a new significant source. There is
more uncertainty about the extragalactic contributions in the submm
and radio.

Although we may forecast wonderfully precise power spectra and cosmic
parameters for the simplest inflation models in
Table~\ref{tab:params}, once we consider the more baroque models with
multifeatured spectra the precision drops(\eg Souradeep \etal
1998). Given that all of our CMB and LSS observations actually access
only a very small region of the inflation potential, imposing
theoretical ``prior'' costs on highly exotic post-inflation shapes
over the observable bands is reasonable. Nonetheless, if the
phenomenology ultimately does teach us that non-baroque inflation and
defect models fail, the CMB and LSS data will be essential for guiding
us to a new theory of fluctuation generation.

We would like to thank George Efstathiou, Lloyd Knox, Dmitry Pogosyan,
and Tarun Souradeep for enjoyable collaborations on a number of the
projects highlighted in the text. 

\section*{References}
\begin{thedemobiblio}{}

\def\prd{{Phys.~Rev.~D}}
\def\prl{{Phys.~Rev.~Lett.}}

\def\name#1{{\it #1\/}}
\def\vol#1{{\bf #1\/}}
\def\jrbfont{\it}
\def\mnras{{\jrbfont Mon. Not. R. Astr. Soc.}}
\def\ApJ{{\jrbfont Astrophys. J.}}
\def\ApJL{{\jrbfont Astrophys. J.~Lett.}}
\def\AstrAst{{\jrbfont Astr. Astrophys.}}
\def\AstrAstSup{{\jrbfont Astr. Astrophys. Suppl. Ser.}}
\def\AstrJ{{\jrbfont Astr. J.}}
\def\apj{{\jrbfont Astrophys. J.}}
\def\apjl{{\jrbfont Astrophys. J.~Lett.}}
\def\apjsuppl{{Astrophys. J.~Supp.}}

\jrbref
Allen, B., Caldwell, R.R., Dodelson,  S., Knox, L., Shellard,
E.P.S. \& Stebbins, A. 1997 preprint astro-ph/9704160.
\jrbref
Beast home page, http://www.deepspace.ucsb.edu/research/Sphome.html
\jrbref
Bennett, C.  \etal 1996a \apjl, {\bf 464}, 1.
\jrbref
Bennett C.  \etal 1996b MAP home page, 
http://map.gsfc.nasa.gov
\jrbref
Bersanelli, M.  \etal 1996 COBRAS/SAMBA, The Phase A Study for an ESA
M3 Mission, ESA Report D/SCI(96)3; 
Planck home page,
http://astro.estec.esa.nl/SA-general/Projects/Cobras/cobras.html
\jrbref
Bond, J.R. 1994 in {\it Relativistic Cosmology},
Proc. 8th Nishinomiya-Yukawa Memorial Symposium, ed. M. Sasaki,
(Universal Academy Press, Tokyo), pp. 23-55.
\jrbref
Bond, J.R. 1996 {\it Theory and Observations of the
Cosmic Background Radiation}, in ``Cosmology and Large Scale
Structure'', Les Houches Session LX, August 1993, ed. R. Schaeffer,
Elsevier Science Press, and references therein.  
\jrbref
Bond, J.R., Efstathiou, G.   \& Tegmark, M.  1997  \mnras\ {\bf 291}, L33 [BET].
\jrbref
Bond, J.R. \&
Jaffe, A. 1997 in {\it Microwave Background Anisotropies},
Proceedings of the XXXI Rencontre de Moriond, ed.  Bouchet, F~R,
Edition Fronti\`{e}res, Paris, pp. 197, astro-ph/9610091.
\jrbref
Bond,~J.R., Jaffe,~A.H.  \& Knox,~L. 1998a \prd\ {\bf 57}, 2117. 
\jrbref
Bond,~J.R., Jaffe,~A.H.  \& Knox,~L. 1998b \apj\ submitted, astro-ph/9808264. 
\jrbref
Bond, J.R.  \& Myers, S. 1996 \apjsuppl\ {\bf 103}, 1-79.
\jrbref 
Bond, J.R., Pogosyan, D. \& Souradeep, T. 1997 ``Proc. 18th
Texas Symposium on Relativistic Astrophysics'', 297-299,
ed. A. Olinto, J.  Frieman and D. Schramm (World Scientific,
Singapore); 1998 Class. Quant. Grav. {\bf 15}, in press; 1998
preprints, CITA-98-22, CITA-98-23.
\jrbref
Boomerang home page, http://astro.caltech.edu/~mc/boom/boom.html
\jrbref
Chaboyer, B., Demarque, P., Krauss, L.M. \&
Kernan, P.J. 1998 \apj\ {\bf 494}, 96. 
\jrbref
Bertschinger, E., Bode, P., Bond, J.R., Coulson, D., Crittenden, R.,
Dodelson, S., Efstathiou, G., Gorski, K., Hu, W., Knox, L., Lithwick,
Y., Scott, D., Seljak, U., Stebbins, A., Steinhardt, P., Stompor, R.,
Souradeep, T., Sugiyama, N., Turok, N., Vittorio, N., White, M.,
Zaldarriaga, M.  1995 ITP workshop on {\it Cosmic Radiation
Backgrounds and the Formation of Galaxies}, Santa Barbara.
\jrbref
Cornish~N.J., Spergel~D.N. \& Starkman~G.D. 1996 {\it
preprint} gr-qc/9602039; 1998 Class. Quant. Grav. {\bf 15}, in press.
\jrbref
Draine~B.T. and  Lazarian~A. 1998
\apjl\ {\bf 494}, 19. 
\jrbref
Efstathiou, G.   \& Bond, J.R. 1998, \mnras\ submitted, astro-ph/9807103. 
\jrbref
Eisenstein, D.J., Hu, W. \& Tegmark, M.  1998 preprint. 
\jrbref
Fixsen, D.J., Cheng~E.S., Gales~J.M., Mather~J.C., 
Shafer~R.A. \& Wright~E.L. 1997 \apj\ {\bf 473}, 576.
\jrbref
Gawiser, E. \& Silk, J.  1998 preprint. 
\jrbref
Gundersen~J.O., Lim~M., Staren~J., Wuensche~C.A.,
 Figueiredo~N., Gaier~T.C., Koch~T., Meinhold~P.R., Seiffert~M.D., 
Cook~G., Segale~A. \& Lubin~P.M. 1995 \apjl\ {\bf 443}, L57-60.
\jrbref
Jaffe, A.H., Knox, L.  \& Bond, J.R. 1997, 
``Proc. 18th Texas Symposium on Relativistic
Astrophysics'', 273-275, ed. A. Olinto, J.  Frieman and D. Schramm (World
Scientific, Singapore), astro-ph/9702109. 
\jrbref
Jungman G.,
Kamionkowski M., Kosowsky A. \& Spergel D.N. 1996 Phys. Rev. D{\bf 54}, 1332.
\jrbref
Knox, L.  1995 \prd\ {\bf 52}, 4307-4318.
\jrbref
Knox, L., Bond, J.R., Jaffe, A.H., Segal, M. \& Charbonneau, D. 1998
preprint. 
\jrbref
Leitch~E.M., Readhead~A.C.S., Pearson~T.J., and Myers~S.T. 1997
\apjl\ {\bf 486}, 23. 
\jrbref
Levin~J.J.,
Barrow~J.D., Bunn~E.F. and Silk~J. 1997 {\it Phys. Rev. Lett.} {\bf 79}
974; Levin~J.J., Scannapieco~E. and Silk~J. 1998
Class. Quant. Grav. {\bf 15}, in press. 
\jrbref
 Lineweaver, C. and Barbosa, D.  1997
astro-ph/9706077. 
\jrbref
 Lineweaver, C. 1998 preprint.
\jrbref
  Hancock, S. and Rocha, G. 1997, astro-ph/9612016, in {\sl
    Proceedings of the XVIth Moriond meeting, ``Microwave Background
    Anisotropies,''} ed.\ F.R.\ Bouchet \etal\ (Gif-Sur-Yvette: Editions
  Fronti\`eres).
\jrbref
MAXIMA home page, http://physics7.berkeley.edu/group/cmb/gen.html
\jrbref
Netterfield, C.B., Devlin, M.J., Jarosik, N., Page, L.  \&
Wollack, E.J. 1997 \apj {\bf 474}, 47.
\jrbref
Peacock~J.A.  1997 \mnras\ {\bf 284}, 885.
\jrbref
Pen, Ue-Li, Seljak, U. \& Turok, N. 1997 preprint astro-ph/9704165.
\jrbref
Perlmutter, S. \etal 1998 preprint. 
\jrbref
Reiss, A.G. \etal 1998 preprint  astro-ph/9805201.
\jrbref
Seljak U. \& Zaldarriaga M. 1996 \apj {\bf 469}, 437.
\jrbref
Souradeep T., Bond~J.R., Knox~L., Efstathiou~G., Turner~M.S. 1998 
astro-ph/9802262. 
\jrbref
Zaldarriaga M.,
Spergel, D. \& Seljak U. 1997 preprint astro-ph/9702157.
\jrbref
White~M., Carlstrom~J.E. and Dragovan~M. 1997, astro-ph/9712195 
\jrbref
TopHat home page, http://cobi.gsfc.nasa.gov/msam-tophat.html

\end{thedemobiblio}{}

\end{document}